\documentclass[reprint,showpacs,showkeys,amsmath,amssymb,superscriptaddress,aps,prl]{revtex4-1}

\usepackage{graphicx}
\usepackage{float}
\usepackage{lipsum}
\usepackage{epstopdf}
\usepackage{bm}
\usepackage{amssymb}
\usepackage{amsmath}
\usepackage{bbold}
\usepackage[colorlinks=true, linkcolor=blue]{hyperref}
\usepackage{ulem}
\usepackage{color, soul}
\usepackage{xcolor}
\usepackage[utf8]{inputenc}
\DeclareUnicodeCharacter{2212}{-}
\DeclareUnicodeCharacter{B0}{\textdegree}
\usepackage{chemformula}
\DeclareUnicodeCharacter{0301}{\'{e}}
\usepackage{gensymb}
\usepackage{graphicx}

\usepackage{tikz}

\begin{document}

\title{Self-induced inverse spin Hall effect in La$_{0.67}$Sr$_{0.33}$MnO$_{3}$ films}

\author {Pushpendra Gupta} 
\address {Laboratory for Nanomagnetism and Magnetic Materials, School of Physical Sciences, National Institute of Science Education and Research (NISER), An OCC of Homi Bhabha National Institute (HBNI), Jatni 752050, India.}

\author {In Jun Park}
\address {Department of Physics, Indiana University – Purdue University Indianapolis, Indianapolis, Indiana 46202, USA}

\author{Anupama Swain}
\address {Laboratory for Nanomagnetism and Magnetic Materials, School of Physical Sciences, National Institute of Science Education and Research (NISER), An OCC of Homi Bhabha National Institute (HBNI), Jatni 752050, India.}

\author{Abhisek Mishra}
\address {Laboratory for Nanomagnetism and Magnetic Materials, School of Physical Sciences, National Institute of Science Education and Research (NISER), An OCC of Homi Bhabha National Institute (HBNI), Jatni 752050, India.}

\author{Vivek P. Amin}
\email{vpamin@iu.edu}
\address {Department of Physics, Indiana University – Purdue University Indianapolis, Indianapolis, Indiana 46202, USA}

\author{Subhankar Bedanta}
\email{sbedanta@niser.ac.in}
\address {Laboratory for Nanomagnetism and Magnetic Materials, School of Physical Sciences, National Institute of Science Education and Research (NISER), An OCC of Homi Bhabha National Institute (HBNI), Jatni 752050, India.}

\address{Center for Interdisciplinary Sciences (CIS), National Institute of Science Education and Research (NISER), An OCC of Homi Bhabha National Institute (HBNI), Jatni 752050, India.}

\date{\today}

\begin{abstract}
The efficient generation of spin currents and spin torques via spin-orbit coupling is an important goal of spintronics research.
One crucial metric for spin current generation is the spin Hall angle, which is the ratio of the spin Hall current to the transversely flowing charge current.
A typical approach to measure the spin Hall angle in nonmagnetic materials is to generate spin currents via spin pumping in an adjacent ferromagnetic layer and measure the transverse voltage from the inverse spin Hall effect in the nonmagnetic layer.
However, given that the spin Hall effect also occurs in ferromagnets, single ferromagnetic layers could generate a self-induced transverse voltage during spin pumping as well.
Here we show that manganite based La$_{0.67}$Sr$_{0.33}$MnO$_{3}$ (LSMO) films deposited by pulsed laser deposition exhibit a significant self-induced inverse spin Hall voltage while undergoing spin pumping.
%
%
A spin pumping voltage of 1.86 $\mu$V is observed in the LSMO (12 nm) film.
Using density functional theory and the Kubo formalism, we calculate the intrinsic spin current conductivities of these films and show that they are in reasonable agreement with the experimental measurements.
\end{abstract}


\maketitle
\section{Introduction}
Many spintronic applications depend on all-electrical control of magnetization dynamics. Such control can be achieved by electrically generating a spin current in a nonmagnetic layer that flows into a neighboring ferromagnetic layer and exerts a spin-orbit torque (SOT) \cite{Liu2012,MihaiMiron2010,RevModPhys.91.035004,Amin2020,9427163}. The spin-transfer torque, or transfer of spin angular momentum to the ferromagnetic layer's magnetization, can result in magnetization switching in magnetic memories and self-sustained oscillations in spin torque oscillators. Thus, an important step in achieving energy-efficient magnetization control is to optimize the efficiency of spin current generation.

It is well known that for nonmagnetic materials with appreciable spin-orbit coupling (such as heavy metals like Pt or Ta), an applied electric field generates a spin current where the electric field, spin flow, and spin polarization are mutually orthogonal. This phenomenon, known as the spin Hall effect \cite{RevModPhys.87.1213}, is typically used as a spin current source in spintronic devices. Likewise, if a spin current with orthogonal spin flow and spin polarization flows into a heavy metal, a potential difference forms that is oriented perpendicularly to the spin flow and spin polarization. This process is known as the inverse spin Hall effect (ISHE). One prominent method to measure the strength of the inverse spin Hall effect in a nonmagnetic material is to generate a spin current via spin pumping \cite{mosendz2010quantifying,hahn2013detection,czeschka2011scaling,wegrowe2011spin} in a neighboring ferromagnetic material and measure the resulting inverse spin Hall voltage.

Both experimental \cite{miao2013inverse,tsukahara2014self,Humphries2017,kanagawa2018self,ASOTWang} and theoretical \cite{PhysRevLett.121.136805,PhysRevResearch.2.033401,mook2020origin,yahagi2021theoretical,PhysRevB.106.024410} studies have shown strong evidence that single ferromagnetic layers also generate transversely-flowing spin currents under an applied electric field. However, due to the lower symmetry of ferromagnets compared to nonmagnets, the electric field, spin flow, and spin polarization need not be mutually orthogonal. Spin current generation in single ferromagnet layers can be attributed to the spin Hall effect \cite{ASOTWang,amin2019intrinsic,omori2019relation}, spin anomalous Hall effect \cite{PhysRevApplied.3.044001,das2017spin,amin2019intrinsic}, spin planar Hall effect \cite{PhysRevApplied.3.044001,PhysRevLett.121.136805,Safranski2018,PhysRevLett.124.197204}, and the magnetic spin Hall effect \cite{davidson2020perspectives,mook2020origin,PhysRevB.106.024410}. It is possible that some of the pumped spin current can flow into the ferromagnetic layer and be converted via the ISHE into a transverse voltage. To confirm this would require measuring an appreciable inverse spin Hall voltage while performing spin pumping in a single ferromagnet layer.

A recent report by Miao et al. shows that the injection of a pure spin current from yttrium iron-garnet (YIG) to NiFe (Py) resulted in an ISHE in Py \cite{miao2013inverse}.
This report suggests the presence of appreciable spin-orbit coupling in the Py film, which is endorsed by the observation of an anisotropic magnetoresistance. Further, Tsukahara et al. \cite{tsukahara2014self} have measured a self-induced ISHE in a single layer Py film, i.e. without any adjacent high spin-orbit coupling material.
These results indicate that the spin current generated within Py could be converted via the ISHE into a tranversely-oriented voltage.
In addition, there are other studies of the intrinsic ISHE in Co and Fe films \cite{kanagawa2018self}.

On a similar note, manganites such as La$_{0.67}$Sr$_{0.33}$MnO$_{3}$ (LSMO), have attracted attention in recent years in spin-to-charge conversion studies due to its low damping, high Curie temperature ($T_{c}$ $\sim$ 350 K), and high spin polarization \cite{luo2017spin,qin2017ultra}.
LSMO is an oxide-based ferromagnet that has been explored only as a spin current source in spin pumping studies \cite{gupta2021simultaneous,luo2017spin,luo2014influence,lee2016magnetic,gupta2023tailoring}.
In particular, in LSMO/Pt bilayers, antidamping like spin torques have been observed \cite{gupta2021simultaneous}, where the spin torques were primarily attributed to the spin Hall effect in Pt. However, to date, a self-induced inverse spin Hall effect has not been reported in manganites like LSMO.

In this work, we report a self-induced ISHE in epitaxial LSMO films under ferromagnetic resonance. Further, we extract the intrinsic spin Hall conductivity of our LSMO films and compare them to first-principles calculations of the bulk spin Hall conductivity. The spin Hall conductivities extracted from both the experimental results and theoretical calculations are on the order of $10~\ohm^{-1}\text{cm}^{-1}$ and thus show reasonable agreement.

\section{Experimental details}
Samples STO(100)/LSMO(t nm) have been deposited via pulsed laser deposition technique having a laser of wavelength 248 nm and a vacuum chamber with base pressure 3 $\times$ 10$^{-7}$ mbar. The samples are called SL1, SL2, and SL3 with thicknesses (t) of 12 nm, 35 nm, and 55 nm respectively. For the growth of LSMO films, the substrate temperature was kept at 740 $\degree$C during the deposition. Laser fluence and frequency are kept at 1.4 J/cm$^2$ and 2 Hz, respectively. The oxygen partial pressure during the growth was maintained at 0.47 mbar. Post-deposition annealing of the sample was performed at the same temperature (740 $\degree$C) for 30 mins at 250 mbar oxygen pressure.
X-ray reflectivity (XRR) and x-ray diffraction (XRD) have been performed with a diffractometer (model- Rigaku smartLab) with wavelength $\lambda$= 0.154 nm.
Cross-sectional high-resolution transmission electron microscopy (HR-TEM) has been performed to check the epitaxy of deposited films by TEM (JEOL, F-200).
The dc magnetometry was performed by a superconducting quantum interference device (SQUID) based magnetometer manufactured by Quantum Design (MPMS3).

Co-planar wave-guide (CPW) based ferromagnetic resonance (FMR) spectroscopy (manufactured by NanOsc, Sweden) was carried out to study magnetization dynamics.
The sample was kept on top of the CPW in a flip-chip manner  \cite{gupta2021simultaneous,singh2019inverse}.
To prevent shunting, a 25 $\mu$m polymer tape was placed between the sample and the CPW.
A dc magnetic field ($H$) was applied perpendicular to the radio frequency field ($h_{rf}$). ISHE measurement was performed at 7 GHz frequency and 25 mW microwave power. The voltage obtained due to ISHE was measured by a Nanovoltmeter (Keithley 2182A).

\section{Results and discussion}

\begin{figure*}[btp]
\centering
 \includegraphics[width=0.85\textwidth]{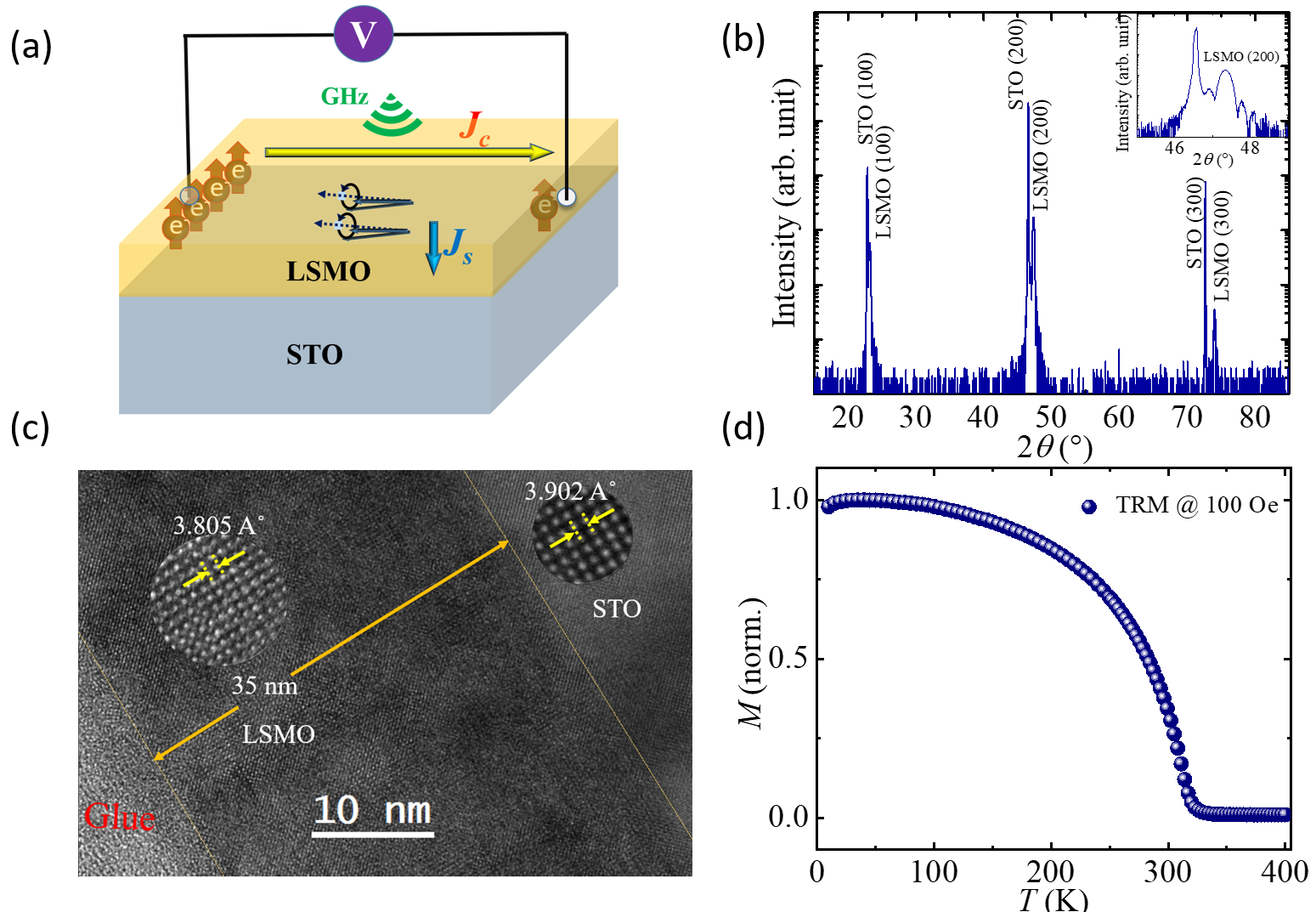}
  \caption{(a) Schematic of sample geometry and ISHE measurement. (b) XRD pattern for sample SL2, in which LSMO film diffraction peaks are observed corresponding to STO (100) planes. The zoomed view of (200) peak is shown in the inset of (b). (c)  HR-TEM image for SL2 sample. The insets are the zoomed part of LSMO films and STO substrate for measuring the lattice constants of LSMO and STO, respectively. (d) Thermoremanent magnetization (TRM) data in which the ferromagnetic to paramagnetic transition is observed at 320 K. }
  \label{fig:figure-1}
\end{figure*}
The schematic of samples studied in this work and ISHE measurement are shown in Fig. \ref{fig:figure-1}(a). The XRD data shown in Fig. \ref{fig:figure-1}(b) convey that LSMO films had been grown in (100) orientations which is also the orientation of the STO substrate.
This confirms that the LSMO films are highly epitaxial in nature.
Film thickness was estimated using X-ray reflectivity (data not shown).
Fig. \ref{fig:figure-1}(c) depicts a cross-sectional HRTEM image of sample SL2, which indicates clear and sharp interfaces of STO/LSMO.
The zoomed-in image confirms the epitaxial nature of LSMO.

\begin{figure}[h]
	\centering
	\includegraphics[width=0.4\textwidth, height=12cm]{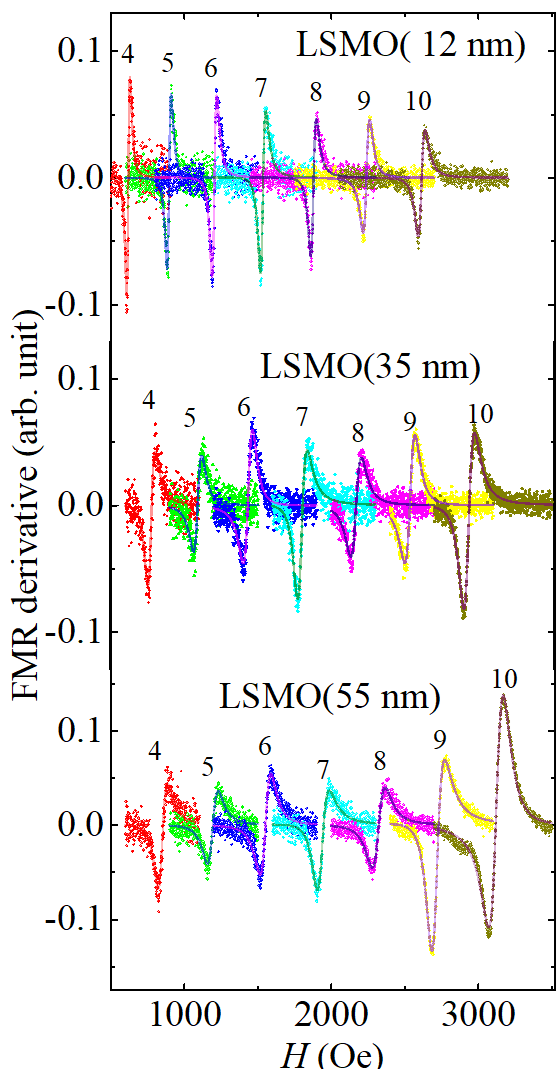}
	\caption{FMR signal (solid circles) for different frequencies for samples SL1, SL2, and SL3. Solid lines are the fits using the Lorentzian equation.}
	\label{fig:figure-2}
\end{figure}

Fig. \ref{fig:figure-1}(d) shows the magnetization (M) vs. temperature (T) plot for the LSMO (35 nm) sample. For this measurement, sample was cooled from 400 K to 5 K in the presence of a 100 Oe magnetic field. After cooling to 5 K, the magnetic field was switched off and magnetization data were recorded with temperature. These protocols are called thermoremanent magnetization (TRM). From the figure, it can be clearly seen that the Curie temperature ($T_{c}$) of the deposited film is 320 K, which confirms the ferromagnetic phase of the sample at room temperature (300 K).

\begin{figure*}[btp]
  \includegraphics[width=0.8\textwidth]{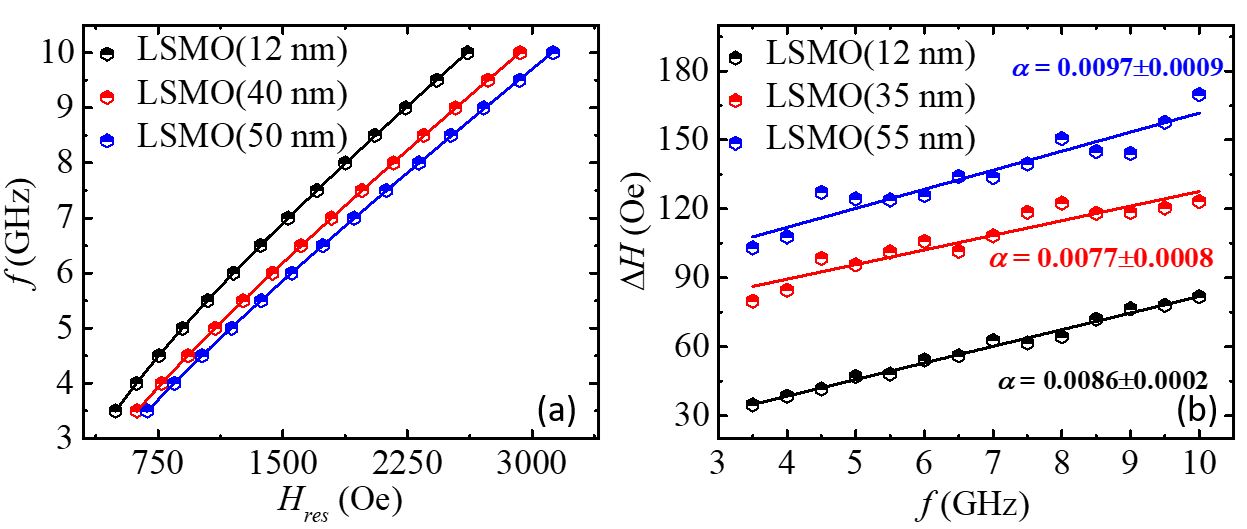}
  \caption{Frequency ($f$) vs. resonance field ($H_{res}$) plot and (b) Linewidth ($\Delta H$) vs. $f$ plot for STO/LSMO samples extracted from Fig. 2. Solid lines are the best fits for equations (1) and (2).}
  \label{fig:figure-3}
\end{figure*}%

To study the magnetization dynamics, we have performed FMR from 3.5-10 GHz. The FMR spectra for all the samples have been shown in Fig. \ref{fig:figure-2}.
To calculate the Gilbert damping constant ($\alpha$), each FMR spectrum was fitted with the derivative of the standard Lorentzian equation.
From the fitting of FMR data, resonance field ($H_{res}$) and linewidth ($\Delta H$) have been extracted.
\textit{f} vs. $H_{res}$ plots are shown in Fig. \ref{fig:figure-3}(a) for all the samples.
The data are fitted using the Kittel equation
\begin{equation}
f=\frac{\gamma}{2\pi}\sqrt{(H_{res}+H_{K})(H_{res}+4\pi M_{eff}+H_{K})}
\label{kittle equation}
\end{equation}

where $\gamma$(=$\frac{g\mu_{B}}{\hbar}$), $g$, $\mu_{B}$, $H_{K}$ and $M_{eff}$ are gyromagnetic ratio, Lande g-factor, Bohr magneton, in-plane anisotropic fields and effective demagnetization, respectively.

Further, \textit{f} vs. $\Delta H$ data shown in Fig. \ref{fig:figure-3}(b) are fitted using the following linear equation 
\begin{equation}
\Delta H = \Delta H_{0}+\dfrac{4\pi\alpha f}{\gamma}
\label{non liner fit}
\end{equation}
where $\Delta H_{0}$ is the inhomogeneous linewidth.
The values of Gilbert damping constant ($\alpha$) for samples SL1, SL2, and SL3 are evaluated to be 0.0086±0.0002, 0.0077±0.0008 and 0.0097±0.0009, respectively.
The values of $\alpha$ match well with the previously reported values \cite{luo2013thickness}. 

\begin{figure}[hb]
	\centering
	\includegraphics[width=0.48\textwidth]{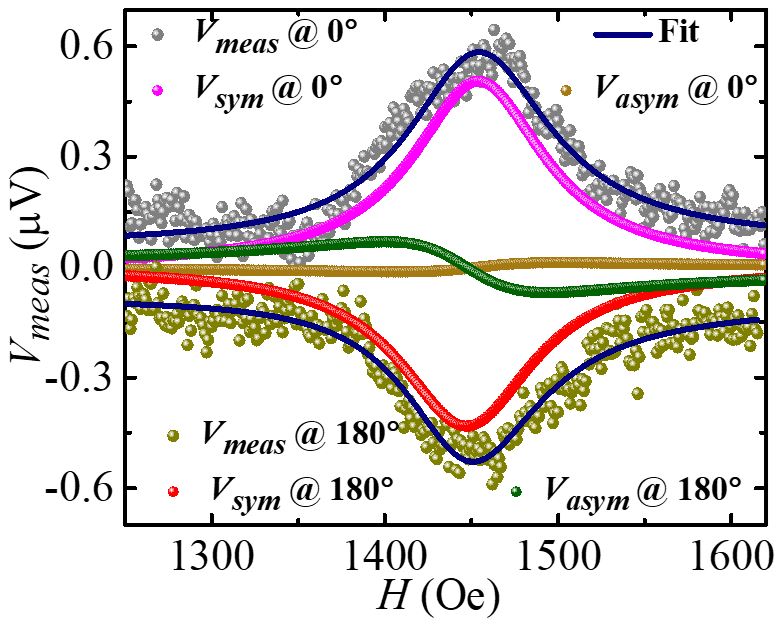}
	\caption{Measured dc voltage ($V_{meas}$) vs. field \textit{(H)} measured at \textit{f}=7 GHz for SL1. $V_{sym}$ and $V_{asym}$ components from $V_{meas}$ have been separated using equation (3).}
	\label{fig:figure-4}
\end{figure}

Further, Gupta et al. has shown the presence of a significant anisotropic magnetoresistance in the LSMO/Pt system, which may arise due to the spin-orbit coupling of the LSMO film (though the adjacent Pt film could contribute via the proximity effect)\cite{gupta2021simultaneous}.
The presence of spin-orbit coupling suggests there could be appreciable spin to charge conversion in the LSMO films.
As LSMO shows metallic ferromagnetic behavior at room temperature, there is a possibility of an intrinsic ISHE in LSMO films similar to that found in the transition metal ferromagnets.
The intrinsic ISHE causes a voltage across the sample transversely oriented to the spin flow, which is measured via making contacts with Cu wires from the sample to a nanovoltmeter.
Fig. \ref{fig:figure-4} shows the measured voltage ($V_{meas}$) data at 0 and 180$\degree$.
$V_{meas}$ fitted with the equation (3) to separate out the symmetric ($V_{sym}$) and asymmetric ($V_{asym}$) contributions to voltage \cite{iguchi2017measurement}.

\begin{equation}
\begin{split}
V_{meas}=V_{sym}\frac{(\Delta H)^{2}}{(H-H_{res})^{2}+(\Delta H)^{2}}+ \\ V_{asym}\frac{(2\Delta H)(H-H_{res})}{(H-H_{res})^{2}+(\Delta H)^{2}}
\end{split}
\end{equation}

From Fig. \ref{fig:figure-4}, it can be clearly seen that the $V_{sym}$ is more dominating than the $V_{asym}$ component.
Also, it can be noticed that $V_{sym}$ changed its polarity when the sample was rotated to 180$\degree$.
This is a characteristic property that confirms the ISHE in LSMO films.

\begin{figure*}
  \includegraphics[width=0.8\textwidth]{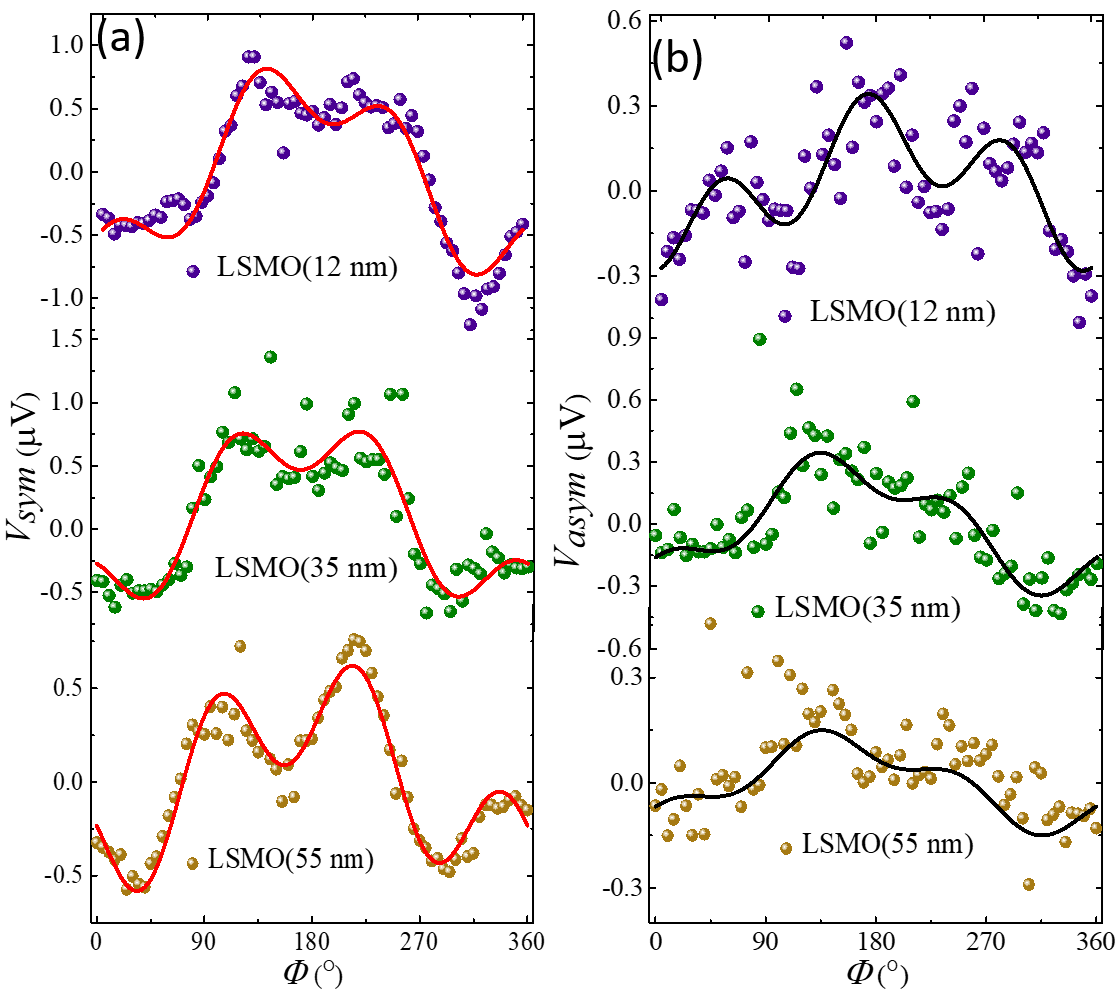}
  \caption{(a) $\phi$ dependent $V_{sym}$ and (b) $V_{asym}$ for LSMO films. Red and black solid lines are the best fits for $V_{sym}$ and $V_{asym}$, using equations (4) and (5), respectively.}
  \label{fig:figure-5}
\end{figure*}

It is well known that the $V_{sym}$ mainly comes due to spin pumping while the spin rectification effects integrated to AMR, AHE, etc., contribute to $V_{asym}$ voltage \cite{iguchi2017measurement}.
To calculate the spin pumping voltage from the measured voltage, a complete angle-dependent ISHE has been performed from 0 to 360$\degree$ in a step of 5$\degree$.
Each measured voltage data is fitted to get $V_{sym}$ and $V_{asym}$ values using equation (3).
Further, $V_{sym}$ and $V_{asym}$ are plotted with their corresponding angles as shown in Fig. \ref{fig:figure-5}(a) and (b).
The $V_{sym}$ data are fitted using the following equation \cite{conca2017lack}
\begin{multline}
V_{sym}=V_{sp} cos^3(\phi + \phi_0)+V_{AHE} cos(\phi + \phi_0) cos\theta+\\ V_{sym}^{AMR\perp} cos(2(\phi + \phi_0)) cos(\phi + \phi_0)+ \\ V_{sym}^{AMR\parallel}  sin(2(\phi + \phi_0)) cos(\phi + \phi_0)
\end{multline}
while the $V_{asym}$ data are fitted using the following equation \cite{conca2017lack}
\begin{multline}
   V_{asym}=V_{AHE} cos(\phi + \phi_0) sin\theta +\\ V_{asym}^{AMR\perp} cos(2(\phi + \phi_0)) cos(\phi + \phi_0) \\ +V_{asym}^{AMR\parallel} sin(2(\phi + \phi_0)) cos(\phi + \phi_0) 
\end{multline}
where $\theta$ is the angle between the electric and magnetic fields of the applied microwave, which is 90$\degree$.
Angle $\phi$ is defined as the angle between the voltage measurement direction and perpendicular to the magnetic field direction, while $\phi_0$ is the extra factor to take care of the misalignment in the sample positioning.
$V_{AHE}$ is anomalous Hall voltage, which arises due to the FM nature of the sample.
$V_{sym}^{AMR \perp,\parallel}$ and $V_{asym}^{AMR \perp,\parallel}$ are perpendicular(parallel) components of symmetric and asymmetric contributions to $V_{AMR}$ and can be calculated using the following equation \cite{conca2017lack}

\begin{equation}
V_{AMR} ^{{\perp,\parallel}}=\sqrt{(V_{asym}^{AMR \perp,\parallel})^{2}+(V_{sym}^{AMR \perp,\parallel})^{2}}
\end{equation}

From the fitting, a significant spin pumping voltage has been obtained for samples, which are listed in Table I.
The value of $V_{sp}$ is dominating over any other spin rectification effects.
As there is no SOC layer adjacent to the LSMO film, $V_{sp}$ can be attributed to the intrinsic ISHE of LSMO films.

\begin{figure}
	\centering
	\includegraphics[width=0.48\textwidth]{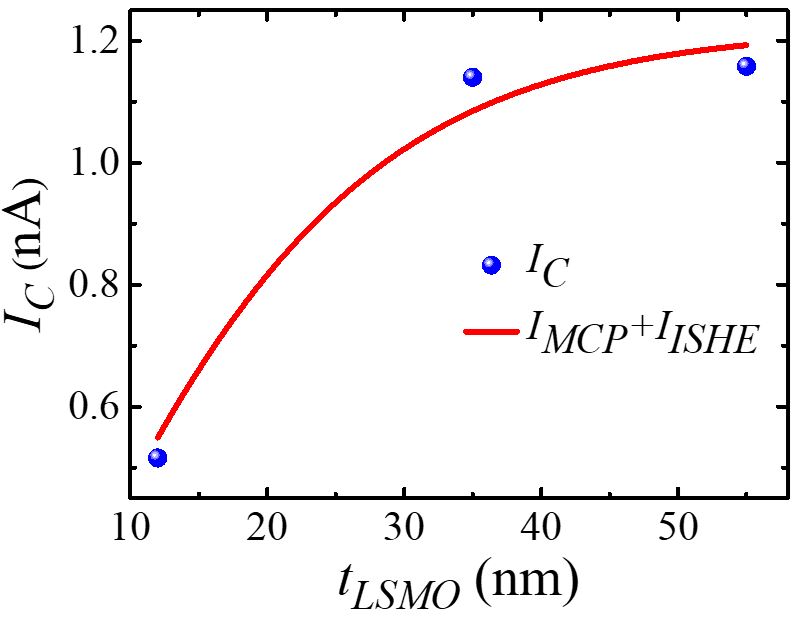}
	\caption{Charge current ($I_{c}$) vs thickness ($t_{LSMO}$) of LSMO. The solid line represents the fitting to $I_{C}=I_{MCP}+I_{ISHE}$ where the $I_{MCP}$ and $I_{SHE}$ are represented by equations (7) and (8), respectively.}
	\label{fig:figure-6}
\end{figure}

Further, $I_{c}$ vs $t_{LSMO}$ data have been plotted as shown in Fig. \ref{fig:figure-6}.
As $I_c$ has a contribution from magnonic charge pumping ($I_{MCP}$) and inverse spin Hall effect ($I_{ISHE}$) hence the data has been fitted using the equation $I_c=I_{MCP}+I_{ISHE}$ where $I_{MCP}$ and $I_{ISHE}$ can be written in terms of following expressions \cite{huang2020spin,azevedo2015electrical}
\begin{multline}
I_{MCP}=(\sigma_{LSMO} t_{LSMO}+\sigma_{STO} t_{STO})w \\ \times (\frac{\Delta \rho}{2M_{s}^{2}})(\frac{M_{eff}\gamma}{4\pi\alpha^{2}}) A^{(d)} h_{rf}^{2}
\end{multline}

\begin{multline}
I_{SHE}=\theta_{SHA} w (\frac{2e}{\hbar})\lambda_{LSMO}\tanh(\frac{t_{LSO}}{2\lambda_{LSMO}})j_{s}^{LSMO}  
\end{multline}
where $\theta_{SHA}$,  $\Delta\rho$, $A^{(d)}$, $h_{rf}$ and $\lambda_{LSMO}$ are the spin Hall angle, anisotropic
magneto-resistivity, spin orbit coupling parameter, RF magnetic field of the microwave, and spin diffusion length, respectively.
$\sigma_{LSMO}$, $t_{LSMO}$, $\sigma_{STO}$, and $t_{STO}$ are the conductivity and thickness of the LSMO and STO layers respectively.
The spin current density at the LSMO/STO interface, $j_{s}^{LSMO}$, is given by \cite{tserkovnyak2005nonlocal,huang2020spin}

\begin{multline}
j_{s}^{LSMO}=\frac{(g_{r}^{\uparrow\downarrow})\gamma^{2} h_{rf}^{2} \hbar [4\pi M_{s}\gamma+\sqrt{(4\pi M_{s}\gamma)^{2}+4\omega^{2}}]}{8\pi \alpha^{2} [(4\pi M_{s}\gamma)^{2}+4\omega^{2}]}
\end{multline}

From the above equation we have calculated the $g_{r}^{\uparrow\downarrow}$ value to 1.2$\times$10$^{16}$ m$^{-2}$and $j_{s}^{LSMO}$ has been considered as a fitting parameter.
From the fitting of the $I_{c}$ vs $t_{LSMO}$ data shown in Fig. \ref{fig:figure-6}, the $\theta_{SHA}$ value is extracted to 0.33.

Finally, we have obtained the spin Hall conductivity using $\theta_{SHA}$=$\sigma_{SH}$⁄$\sigma_{C}$ spin Hall conductivity and charge conductivity, respectively.
The calculated values of $\sigma_{SH}$ for SL1, SL2, and SL3 are 76.4, 58.9, and 50.0  $\ohm^{-1}cm^{-1}$, respectively.

In an earlier report spin current has been generated in the Py layer itself via FMR due to the high resistivity of the adjacent SiO$_{2}$ layer which prevents the spins to diffuse through it (SiO$_{2}$  layer) \cite{tsukahara2014self}.
A similar self-induced ISHE observation has also been made in Co and Fe films \cite{kanagawa2018self}.
In our case, the resistivity of STO is comparatively less than SiO$_{2}$  \cite{saadatkia2018photoconductivity}.
When FMR occurs in the LSMO layer, the spin current is generated, which tends to move towards the LSMO/STO interface.
%
%
At the interface, spins get scattered due to the charged impurities which might be causing a spin gradient in the LSMO layer.
This spin gradient creates spin current which gets converted into charge current via ISHE, which leads to the generation of potential difference.

\begin{table}
\caption{$V_{sp}$, and $V_{AHE}$ values for all samples from the fitting of the data shown in Fig. \ref{fig:figure-5}(a) and (b) using equations (4) and (5). $V_{AMR}$ values were calculated using equation (6).} 
\centering
\resizebox{0.35\textwidth}{!}
{
\begin{tabular}{l l l l l}
\hline
Sample & $V_{sp}$ & $V_{AHE}$ & $V_{AMR} ^{\perp}$ & $V_{AMR} ^{\parallel}$\\
 & ($\mu V$) & ($\mu V$) & ($\mu V$) &($\mu V$)\\
\hline
S1 & 1.86 & 0.12 & 1.43 & 0.39\\    
\hline
S2 & 1.83 & 0.31 & 1.47 & 0.13\\    
\hline
S3 & 1.39 & 0.12 & 1.31 & 0.15\\    
\hline
\end{tabular}
}
\end{table}

\section{Computation Details}

\begin{figure}[h]
	\centering
	\includegraphics[width=0.3\textwidth]{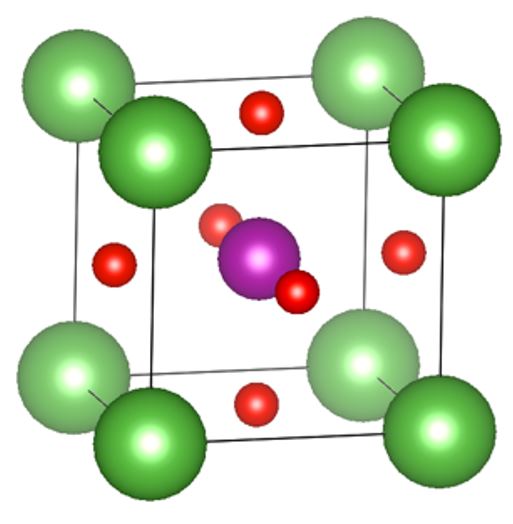}
	\caption{Bulk crystal structure of perovskite LaMnO$_3$ or SrMnO$_3$. The green, red, and purple spheres represent La (or Sr), O, and Mn atoms, respectively. A simpler model was chosen to reduce computational costs.}
	\label{fig:figure-7}
\end{figure}

In order to evaluate the appreciable ISHE in LSMO from the experimental work, we performed the density functional theory (DFT) calculations as implemented in the Quantum Espresso package\cite{giannozzi2017advanced,giannozzi2009quantum,giannozzi2020quantum}.
To reduce computational costs, we limited our theoretical calculation to LaMnO$_{3}$ (LMO) and SrMnO$_{3}$ (SMO), instead of LSMO since LSMO requires a large unit cell in order to meet its chemical composition with symmetry and redundant calculations to average the disorder.
The representative unit cell of LMO or SMO is shown in Fig. \ref{fig:figure-7}, where green, red, and purple spheres are La (or Sr), O, and Mn atoms, respectively.
Cut-off energies for plane wave and charge density of 90 Ry and 900 Ry and 9$\times$9$\times$9 k-points grids were used to ensure the total energy converged within 7$\times$10$^{-9}$ Ry.
Once the ground states were obtained, we interpolated them into an orbital basis using the Wannier90 package \cite{pizzi2020wannier90}.
With interpolated basis, the spin Hall conductivity tensor was calculated using Kubo’s formula as implemented in \cite{qiao2018calculation,tsirkin2021high}
\begin{multline}
\sigma_{\alpha\beta\gamma}=\frac{-e\hbar}{N_{k}V_{c}} \sum_{k}\sum_{n,m}(f_{nk}-f_{mk}) \times \\ \frac{Im[\langle \psi_{nk}|\frac{1}{2}\{S^{\gamma},\nu_{\alpha}\}|\psi_{mk}\rangle \langle \psi_{mk}|\nu_{\beta}|\psi_{nk}]}{(\epsilon_{nm}-\epsilon_{mk})^2-(\hbar \omega)+i\eta)^2}
\end{multline}
where $\alpha$, $\beta$, and $\gamma$ are the direction of spin current, electric field, and spin polarization, respectively.
400$\times$400$\times$400 k-grids were used to calculate the spin Hall conductivity tensors of LMO and SMO.

For non-magnetic materials, it is known that the non-zero elements of spin Hall conductivity tensor can occur when the three components (i.e., $\alpha$, $\beta$, and $\gamma$) are mutually orthogonal, and they have the same magnitude.
However, due to the lower symmetry from magnetization, ferromagnetic materials have different non-zero spin Hall conductivity components depending on the magnetization direction \cite{davidson2020perspectives}.
For instance, when the magnetization is parallel to $x$ direction, $\sigma_{xyz}=-\sigma_{xzy}$, $\sigma_{yzx}=-\sigma_{zyx}$, and $\sigma_{zxy}=-\sigma_{yxz}$ while it is parallel to $z$ direction, $\sigma_{xyz}=-\sigma_{yxz}$, $\sigma_{yzx}=-\sigma_{xzy}$, and $\sigma_{zxy}=-\sigma_{zyx}$.
Therefore, we showed our calculation results for only distinctive non-zero components (i.e., $\sigma_{xyz}$, $\sigma_{yzx}$, and $\sigma_{zxy}$), as shown in Table II.
Clearly, there are non-zero values for spin Hall conductivity tensor on the order of 10 $\ohm^{-1}cm^{-1}$, which are on the same order from experimentally obtained values, although they are smaller than the ones from other perovskite structures ($\sim$ 100 $\ohm^{-1}cm^{-1}$ ), such as SrIrO$_3$ \cite{nan2019anisotropic}, SrRuO$_3$ \cite{ou2019exceptionally}, BaOsO$_3$, and SrOsO$_3$ \cite{jadaun2020rational} or Pt ($\sim$ 2000 $\ohm^{-1}cm^{-1}$) \cite{qiao2018calculation}. 
We note that extrinsic mechanisms of the ISHE could also contribute to the experimental measurements but are not included in the theoretical calculations.

\begin{table}
\caption{Calculated non-zero spin Hall conductivity tensor elements of LMO and SMO} 
\centering
\resizebox{0.5\textwidth}{!}
{
\begin{tabular}{l l l l}
\hline
Spin Hall conductivity & $\sigma_{xyz}$ & $\sigma_{yzx}$ & $\sigma_{zxy}$ \\
 & ($\ohm^{-1}cm^{-1}$) & ($\ohm^{-1}cm^{-1}$) & ($\ohm^{-1}cm^{-1}$)\\
\hline
$LaMnO_{3} (M\parallel x)$ & 17.457 & -18.671 & 14.547\\    
\hline
$LaMnO_{3} (M\parallel z)$ & -18.677 & 14.558 & 14.471\\    
\hline
$SrMnO_{3} (M\parallel x)$ & 1.066 & 16.657 & -3.595\\    
\hline
$SrMnO_{3} (M\parallel z)$ & 17.050 & -3.638 & -1.019\\    
\hline
\end{tabular}
}
\end{table}

\section{Conclusion}
In summary, we measured a self-induced ISHE in highly epitaxial LSMO films and performed first-principles calculations of the spin Hall conductivities of bulk LSMO.
The LSMO films were prepared using pulsed layer deposition and characterized by different techniques to confirm their epitaxial nature.
The spin pumping voltage ($V_{sp}$) dominates over any other spin rectification effects in our samples, indicating the presence of appreciable spin-to-charge conversion in LSMO films.
The spin Hall conductivity was calculated from first principles using density functional theory and the Kubo formalism, and the results qualitatively match our experimental findings.
This study will help shed light on spin current generation in manganite-based films.

\section{Acknowledgments}
SB, PG, AM and AS thank the Department of Atomic Energy (DAE) and SERB (Science and Engineering Research Board), Govt. of India, for providing financial support.  PG acknowledges UGC for SRF fellowship. VA and IJP acknowledge support from the National Science Foundation (NSF DMR-2105219).
\bibliography{References} 
\end{document}